# $d^2$, $d^3$, and $d^4$ M$_3$C$_2$ Transition Metal Carbides (MXenes) as Catalysts for CO$_2$ Conversion into Hydrocarbon Fuels: A Mechanistic and Predictive DFT Study

Neng Li,[a,‡] Luis Miguel Azofra,[b,‡] Xingzhu Chen,[a] Douglas R. MacFarlane,[b] Xiujian Zhao,[a] and Chenghua Sun[b,*]

[a]State Key Laboratory of Silicate Materials for Architectures, Wuhan University of Technology, Hubei, 430070, China
[b]ARC Centre of Excellence for Electromaterials Science (ACES), School of Chemistry, Faculty of Science, Monash University, Clayton, VIC 3800, Australia

**ABSTRACT:** The functioning of 2D $d^2$, $d^3$, and $d^4$ transition metal carbides (MXenes) with formulae M$_{n+1}$C$_n$ ($n$ = 2) as CO$_2$ conversion catalysts has been proved by well-resolved density functional theory (DFT) and DFT+$U$ theoretical calculations. Whilst MXenes from the $d^2$ series (M = Ti, Zr, and Hf) have demonstrated active behaviors for the capture of CO$_2$, the V$_3$C$_2$, Nb$_3$C$_2$, Cr$_3$C$_2$, and Mo$_3$C$_2$ materials exhibit the most promising results for their application in the selective CO$_2$ conversion into CH$_4$, with limiting reaction energies of 1.55, 1.75, 0.69, and 1.24 eV, respectively, at DFT+$U$ computational level plus explicit DFT-D3 dispersion corrections, and specially highlighting the role of Cr$_3$C$_2$ due to its theoretically predicted low over-potential. Moreover, important features have been predicted during the first hydrogenation step towards the formation of the OCHO• and HOCO• radical species, exhibiting spontaneous reaction energies in the OCHO• obtaining for such promising carbides from the $d^3$ and $d^4$ groups. Our results provide novel insights in the computer-aided searching of high performance catalysts and the understanding of reaction mechanisms for CO$_2$. Finally, it is hypothesized that the capture of CO$_2$ along the early step of the reaction is spontaneously produced without pass from a physisorbed state, being the strength of such capture larger than the computed binding energies for the chemisorption of H$_2$O, offering encouraging perspectives for the experimental testing of our materials in water environment.

## INTRODUCTION

The massive and large-scale anthropogenic carbon dioxide (CO$_2$) emissions into the atmosphere, as a consequence of our heavy reliance on imported non-renewable energy sources,[1] have triggered the intensification of the greenhouse effect, a serious environmental problem which has been related to climate change.[2] The magnitude of this issue confronts us with an unprecedented challenge: significant reductions in CO$_2$ emissions and the search for realistic alternatives to the energy sources based on petrol are critical.

In this regard, CO$_2$ conversion technology,[3] as an energy storage alternative due to its relevance for the environment and sustainability, has been actively developed towards the production of hydrocarbon 'green fuels' that can be re-burned for energy generation with a zero-balance of greenhouse emissions. Since the CO$_2$ conversion technology is profiled as one of the several solutions to the so-called 'CO$_2$ problem', a vast and rich literature has emerged in such topic in the recent years: a diverse selection of materials have been extensively used and optimized such as titania-based (TiO$_2$) semiconductors,[4-6] Cu and photo-catalytic active Cu$_2$O,[7-9] 2D metal-free meshes as functionalized graphene oxide[10] or graphite-like carbon nitride (g-C$_3$N$_4$),[11] among others.[12-15] However and as indicated by Whipple *et al.*,[16] many advances in the field are still needed to address the energy efficiency and the reaction rate issues, among others limitations.

From a mechanistic point of view, the CO$_2$ conversion process through electrochemical approaches consists in successive electro-reductions by inclusion of a set of H$^+$/e$^-$ pairs, and depending on the total even number of H$^+$/e$^-$ couples transferred along the whole reaction (from two to eight), different hydrocarbon compounds such as carbon monoxide (CO) or formic acid (HCOOH), formaldehyde (H$_2$CO), methanol (CH$_3$OH), or methane (CH$_4$), can be obtained.[17] In this regard, an advanced understanding of the reaction mechanism is essential for the design of novel catalysts, specially when specific products are targeted. On the one hand, catalysis cannot occur if there is no effective physicochemical contact between gas molecules and the catalyst surface. A non-covalent interaction, normally of O=C=O$_{lp}$···Y nature with "Y" being an electro-positive atom, is required in order to adsorb CO$_2$ on the surface. However, CO$_2$ fixa-tion is often thermodynamically non-spontaneous under mild conditions (i.e. ambient temperatures and pressures) and additional energy/pressure is required to enhance such contact. Once it is adsorbed, the next challenge is the first hydrogenation step, denoted by the single-electron $CO_2 + e^- \rightarrow CO_2\bullet^-$ process. This represents a large impediment for the overall process since a significant amount of energy is required. In that regard, a strongly negative reduction potential of $-1.90$ V vs. NHE is exhibited,[18] constituting the limiting step of the whole electrochemical reaction even when catalysts come into play. In addition to that, reduction potentials vs. NHE at neutral pH indicate a very poor selectivity for the isolated transformations into the aforementioned potential hydrocarbon products. To overcome these issues, the finding of novel strategies, specially new catalysts, is required.

Recent work carried out in our group have provided novel insights along the early stages of the $CO_2$ conversion mechanism when doping two dimensional (2D) boron nitride nano-sheets or meshes (BNs) with beryllium.[19] As the consequence of the insertion of such electron-deficient atoms to the 2D network, important diminutions in the reaction energies were demonstrated for the $CO_2$ fixation and the first hydrogenation steps, i.e. due to the very deep π-hole generated on the Be-doped surface environment, the $CO_2$ adsorption became spontaneous in terms of the Gibbs free binding energy. More remarkably, the radical HOCO• and OCHO• species formed as result of the first $H^+/e^-$ gain also exhibited spontaneous thermodynamics, with Gibbs free reaction energies at room temperature being equal to $-0.45$ and $-0.98$ eV, respectively. For classical semiconductors, the first proton-electron transfer often constitutes the limiting step of the whole reaction,[18, 20, 21] while the outcomes generated through our DFT results indicate that this can be effectively changed through introducing special active sites on the surface.

However, despite the significance of such insights and the theoretical value from the computer-aided design of catalysts perspective, the difficulties involving Be-doping of BNs, as well as the high toxicity of beryllium,[22] limit its potential applicability. In this context and following similar features and mechanisms, transition metal carbides emerge as promising candidates for such purposes. Recent investigations carried out by Naguib et al.[23] have demonstrated that the exfoliation of strong primary bond containing solids, such as MAX phase powders of $Ti_3AlC_2$ into $Ti_3C_2$, labeling this novel kind of 2D materials as 'MXenes' due to their similarity with graphene. As has been recently summarized by Gogotsi and co-workers,[24] the etching out of the "A" layers from MAX phases with formulae $M_{n+1}AX_n$, which "M" an early transition metal, "A" an atom from the triel or tetrel groups (i.e. icosagens and crystallogens, respectively), and "X" = C or N with n = 1, 2, and 3, produces the synthesis of their respective MXenes with formulae $M_{n+1}X_n$ prior applying sonication technique. In this regard, not only $Ti_3C_2$ has been synthesized via this procedure, but also $Ti_2C$, $Nb_2C$, $V_2C$, $(Ti_{0.5}Nb_{0.5})_2C$, $(V_{0.5}Cr_{0.5})C_2$, $(Nb_{0.8}Ti_{0.2})_4C_3$, $(Nb_{0.8}Zr_{0.2})_4C_3$, $Ti_3CN$, $Ta_4C_3$, $Mo_2TiC_2$, $Mo_2Ti_2C_3$, or $Cr_2TiC_2$, among others,[23, 25, 26-28] which with large specific surface area are typically needed to obtain large capacitances for further $CO_2$ capture applications.

MXenes, such 2D solids,[24] have generated recent interest because of their outstanding electronic properties which can be exploited for innumerable industrial and biomedical applications. Multilayer MXenes are conductively similar to multilayer graphene. However, unlike graphene, MXenes can be easily dispersed in aqueous solutions because of their hydrophilic properties. Also, the inherent metallic character of MXenes may be modified to act as semiconductor once the surface is –F or –OH terminated.[25]

In term of real applications, MXenes have already been demonstrated to be promising candidates for energy storage applications such as Li-ion batteries,[29-31] non-Li-ion batteries,[32] electrochemical supercapacitors,[33-35] and fuel cells.[36] Apart from the energy storage applications, MXenes were tested as photocatalytic materials,[37] gas sensors,[38] biosensors,[39] and transparent, conductive electrodes.[40]

In the recent years, some literature has emerged describing the applicability of metal carbides as catalytic substrate for hydroprocessing, water splitting technology,[41-44] given the inherent properties of these materials with high specific areas and cleanness surfaces,[45] good electrical conductivities, stability, and hydrophilic behaviors.[25] Based on this background, we hypothesize and report theoretical evidences highlighting the function of $d^2$, $d^3$, and $d^4$ MXenes, transition metal carbides with formulae $M_{n+1}C_n$ (n = 2) as $CO_2$ capture and conversion catalysts. In this regard, our DFT and state-of-the-art DFT+$U$ studies plus dispersion corrections predict active behaviors for the capture of $CO_2$, being the strength of such capture larger than the computed binding energies for the chemisorption of $H_2O$. These inherent features expect an active physics for the $CO_2$ conversion mechanism: while injection of $CO_2$ solute is spontaneously produced on the material, the hydrophilic and high electrical conductivity advert the effective electro-reductions in both the $H^+$ attachments and electrons providing.

## COMPUTATIONAL DETAILS

The mechanism for the electrochemical conversion of $CO_2$ into hydrocarbon compounds catalyzed by $d^2$ (Ti, Zr, and Hf), $d^3$ (V, Nb, and Ta), and $d^4$ (Cr and Mo) 2D transition metal carbides or MXenes, with formulae $M_{n+1}C_n$ (n = 2), has been studied by means of density functional theory (DFT) through the generalized gradient approximation (GGA) with the Perdew-Burke-

Ernzerhof (PBE) functional,[46] using a plane-wave cut-off energy of 400 eV.[47, 48] Concerning the periodic boundary conditions, the Brillouin zone was sampled by 3×3×1 k-points using the Monkhorst-Pack scheme, after the tests with a larger set of k-points to make sure that there was no significant changes in the calculated energies. In order to avoid interactions between periodic images, a vacuum distance of 20 Å was imposed between different layers. At first stage, full optimizations were carried out using energy and force convergence limits equal to $10^{-4}$ eV/atom and |0.05| eV/Å, respectively. For energy calculations, van der Waals interactions were taken into account through the Grimme DFT-D2 method.[49] In order to evaluate the zero point energy (ZPE) as well as the thermal corrections terms, additional calculations over the Γ points were carried out.

After the first-round calculations based on gas-phase models, the $d^3$ $V_3C_2$ and $Nb_3C_2$, and the $d^4$ $Cr_3C_2$ and $Mo_3C_2$ surfaces have identified as the most promising catalysts for the $CO_2$ conversion. To further examine such results, additional Hubbard-like parameters have been included through the DFT+$U$ approach in the form developed by Dudarev et al.,[50, 51] to treat the strong on-site Coulomb interactions of the $d$-like localized electrons in $M_3C_2$, which are not correctly described via classical GGA-based DFT. In our case, $U$ terms equal to 3.1, 3.0, 3.5, and 3.5 eV have been employed for V, Nb, Cr, and Mo, respectively.[52-55] Over such DFT+$U$ re-optimizations, explicit dispersion correction terms to the energy were employed through the use of the DFT-D3 method with the standard parameters programmed by Grimme and co-workers.[56, 57] With the aim to obtain more accurate values, computational settings were modified using a plane-wave cut-off energy of 450 eV, |0.02| eV/Å as force convergence limit, and the Brillouin zone was expanded by 5×5×1 k-points. All optimization calculations have been performed through the facilities provided by the Vienna Ab-Initio Simulation Package (VASP, version 5.3.5).[58-61]

Finally, Eqn. (1) has been applied to calculate the reaction energies, where $n$ is the number of $H^+/e^-$ pairs transferred, $m$ the number of $H_2O$ molecules released, and $p$ the number of $CH_4$ molecules released, if applicable. In such context, Nørskov and co-workers[20] estimate that the chemical potential of the $H^+/e^-$ pair has the half value of the chemical potential of the dihydrogen ($H_2$) molecule [see Eqn. (2)].

$$\Delta G_R = G(\text{surf} \cdots CO_{2-m-p}H_{n-2m-4p}) + m\,G(H_2O) + p\,G(CH_4) - G(\text{surf}) - G(CO_2) - n/2\,G(H_2) \quad (1)$$

$$\mu(H^+/e^-) = \tfrac{1}{2}\mu(H_2) \quad (2)$$

## RESULTS AND DISCUSSION

Transition metal carbides (MXenes) with formulae $M_{n+1}C_n$ ($n = 2$) and M = Ti, Zr, Hf, V, Nb, Ta, Cr, and Mo (hereafter simply referred as $M_3C_2$), are strongly bounded graphene-similar 2D materials composed by five layers of atoms in which carbons are in the inner layers being six-fold coordinated (labeled as 6c-C in **Fig. 1**) with two kinds of transition metals: those constituting the central inner layer (also hexa-coordinated) and the three-fold terminated (labeled as 3c-M in **Fig. 1**) which are specially reactive due to their empty $d$-like orbitals, and therefore where the catalytic activity will take place.

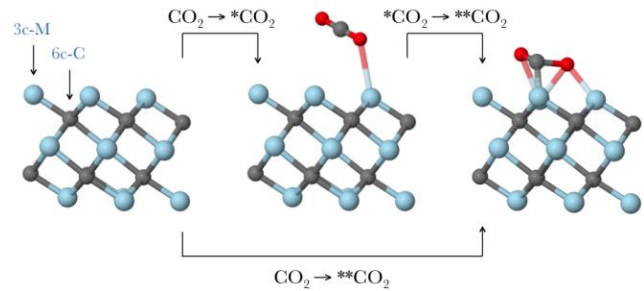

**Figure 1.** Proposed paths for the $CO_2$ interaction with the $M_3C_2$ MXenes surfaces: clean surface (left), $CO_2$ physisorption (center), and $CO_2$ chemisorption (right). 3c-M and 6c-C labels refer to three-fold transition metal and six-fold carbon atoms, respectively.

As indicated in **Fig. 1**, the interaction between $CO_2$ and the M-terminated surface of the MXenes might be carried out either through a physical fixation (physisorption) where $CO_2$ is attached to the surface through a non-covalent interaction of O=C=O$_{lp}$···M nature (**Fig. 1**, center), or through a chemisorbed process in which $CO_2$ is formally bound (**Fig 1**, right). Gibbs free binding energies at room temperature ($\Delta G_b$) display clear differences in respect to the thermodynamic stability in each case. On the one hand, **Table 1**, the $CO_2$ physisorption step is characterized by non-spontaneous binding energies, between 0.15 and 0.35 eV. This is in contrast to the case of $Ta_3C_2$ (see **Table 1**) as catalyst in which no minimum was located and $Ti_3C_2$ where the physical $CO_2$ fixation is spontaneously produced with –0.59 eV. On the other hand, the MXenes seem to be active towards $CO_2$ chemisorption, exhibiting spontaneous binding energies decreasing in strength as we move along the group, i.e. surfaces for $d^2$ MXenes are stronger capture materials than $d^3$ and $d^3$ stronger than $d^4$. At this point, the question arising is: does the process of capture occur directly or is it done in two stages after passing the barrier imposed by the $CO_2$ physisorption?

In this regard, since MXenes are highly reactive surfaces, it seems contradictory to find such high barriers for the $CO_2$ fixation step, with values comparable or even larger than those observed with less-reactive surface materials, such as rutile $TiO_2(110)$.[62] Comparing both classical DFT test results including van der Waals interactions through the Grimme's DFT-D2 method and state-of-the-art and more accurate DFT+$U$ calculations with explicit dispersion corrections *via* the DFT-D3 method (see **Tables 1** and **2**), it seems that as consequence of the force convergence settings (see **Computational details** section), physisorbed $CO_2$ minima appear as DFT artifacts, allowing us to conclude that $CO_2$ directly interacts with the surface through a spontaneous and exothermic process that produces its capture.

**Table 1.** Gibbs Free Reaction Energies (in eV), calculated at PBE (GGA-DFT) computational Level (Thermal and ZPE Corrections included). All values are commonly referred to the Clean Surface and the Isolated Reactive Gases (0.00 eV). Note: * and ** Symbols Refer to Physisorbed and Chemisorbed Species, respectively.

|  | $Ti_3C_2$ | $Zr_3C_2$ | $Hf_3C_2$ | $V_3C_2$ | $Nb_3C_2$ | $Ta_3C_2$ | $Cr_3C_2$ | $Mo_3C_2$ |
|---|---|---|---|---|---|---|---|---|
| $*CO_2$ | –0.59 | 0.17 | 0.18 | 0.29 | 0.35 | — | 0.25 | 0.15 |
| $**CO_2$ | –2.99 | –3.16 | –3.03 | –1.45 | –1.61 | –2.29 | –1.28 | –2.12 |
| $**OCHO\bullet$ | –2.05 | –2.25 | –2.88 | –1.38 | –1.72 | –1.59 | –1.60 | –1.77 |
| $**HOCO\bullet$ | –2.07 | –2.49 | –2.79 | –1.40 | –1.56 | –1.94 | –1.74 | –1.94 |
| $**\bullet OCH_2O\bullet$ | –3.53 | –4.09 | –4.32 | –1.93 | –2.25 | –2.88 | –1.62 | –1.69 |
| $**HCOOH$ | –1.02 | –2.20 | –2.48 | –0.15 | –0.15 | –0.35 | –0.01 | –0.83 |
| $**CO$ | –1.27 | –1.19 | –1.61 | –1.53 | –1.49 | –1.90 | –2.09 | –2.39 |
| $**HOCH_2O\bullet$ | –2.50 | –2.84 | –3.13 | –1.60 | –1.92 | –2.59 | –1.88 | –2.21 |
| $**HOCH_2OH$ | –1.14 | –1.10 | –1.55 | –0.64 | –0.73 | –1.15 | –0.73 | –0.97 |
| $**H_2CO$ | –2.55 | –3.31 | –3.41 | –1.91 | –2.28 | –2.50 | –1.89 | –2.00 |
| $*H_2CO$ | 1.05 | 0.94 | 0.89 | 0.91 | 0.96 | 0.85 | 0.96 | 0.89 |
| $**CH_2OH\bullet$ | –1.71 | –2.17 | –2.00 | –1.37 | –1.49 | –2.07 | –1.63 | –1.79 |
| $**CH_3O\bullet$ | –3.06 | –3.23 | –3.44 | –2.31 | –2.49 | –3.11 | –2.24 | –2.68 |
| $**CH_2$ | –2.01 | –2.08 | –1.38 | –1.16 | –1.69 | –2.54 | –1.86 | –2.34 |
| $*CH_3OH$ | –0.08 | –0.05 | –0.12 | 0.21 | –0.06 | 0.08 | 0.20 | 0.01 |
| $**CH_3\bullet$ | –2.25 | –2.58 | –3.07 | –2.46 | –2.70 | –3.49 | –2.73 | –3.23 |
| $*CH_4$ | -1.41 | –0.91 | –1.15 | –2.36 | –1.00 | –1.06 | –0.78 | –0.96 |
| $**O\cdots*CH_4$ | –3.97 | –4.30 | –4.48 | –2.59 | –3.13 | –3.44 | –2.56 | –2.64 |
| $**OH\bullet$ | –4.09 | –4.14 | –4.29 | –3.19 | –3.41 | –3.85 | –3.23 | –3.44 |
| $**H_2O$ | –2.54 | –2.47 | –2.54 | –1.96 | –1.89 | –2.16 | –2.07 | –2.42 |

More remarkable are the huge differences shown between the Gibbs free binding energies computed at PBE/DFT-D2 and DFT+$U$/DFT-D3 levels of theory. In this regard, the strong on-site Coulomb interactions of the *d*-like localized electrons in $M_3C_2$ treated through the inclusion of explicit $U$ Hubbard-like parameters, seems to be essential in the adequate description of the captured $CO_2$ steps: comparisons for $\Delta G_b$ in the $d^3$ $V_3C_2$ and $Nb_3C_2$, and the $d^4$ $Cr_3C_2$ and $Mo_3C_2$ materials differ in around 1.0-1.3 eV, supporting the spontaneity of the chemisorption process, but correcting the errors derived from over-estimated interactions between $CO_2$ and the M-terminated MXene surfaces.

The outcomes presented in this work are of great significance since they suggest that $d^2$-$d^4$ $M_3C_2$ MXenes are capable of spontaneously producing the capture of $CO_2$, overcoming the mechanistic limitation imposed by this first step of the $CO_2$ fixation, that as has been commented before, usually demands the input of energy/pressure to enhance the contact on the surface.[63]

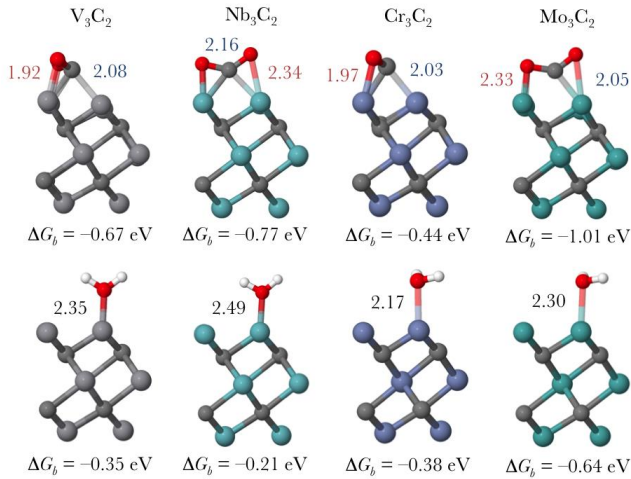

**Figure 2.** At top, captured $CO_2$ minima. Selected distances in dark blue and red indicate the proximal C–M and O–M distances, respectively, in Å. At bottom, $H_2O$ chemisorption steps with proximal O–M distances in Å. Gibbs free binding energies at room temperature calculated at DFT+$U$/DFT-D3 computational level, with $U$ = 3.1, 3.0, 3.5, and 3.5 eV for V, Nb, Cr, and Mo, in their respective $M_3C_2$ MXenes.

**Table 2.** Gibbs Free Reaction Energies (in eV), calculated at DFT+$U$/DFT-D3 (See **Computational Details**) Computational Level (Thermal and ZPE Corrections included) for $d^3$ $V_3C_2$ and $Nb_3C_2$, and $d^4$ and $Cr_3C_2$ and $Mo_3C_2$ materials. All values are referred to the clean surface and the isolated reactive gases (0.00 eV). Note: * and ** Symbols Refer to Physisorbed and Chemisorbed Species, respectively.

|  | $V_3C_2$ | $Nb_3C_2$ | $Cr_3C_2$ | $Mo_3C_2$ |
|---|---|---|---|---|
| **$CO_2$ | –0.77 | –0.67 | –0.44 | –1.01 |
| **OCHO• | –1.32 | –1.06 | –1.12 | –1.42 |
| ***•OCH_2O• | –1.38 | –2.33 | –0.83 | –1.51 |
| **HOCH_2O• | –1.41 | –1.53 | –0.94 | –1.80 |
| **$H_2CO$ | –1.75 | –1.84 | –0.82 | –1.74 |
| **$CH_3O$• | –1.94 | –2.47 | –1.32 | –2.18 |
| **O ···*$CH_4$ | –1.99 | –2.95 | –1.23 | –1.83 |
| **OH• | –2.96 | –3.30 | –2.26 | –3.08 |
| **$H_2O$ | –1.41 | –1.55 | –1.58 | –1.84 |

Additionally, and as a paramount feature, DFT+$U$/DFT-D3 calculations of the $d^3$ $V_3C_2$ and $Nb_3C_2$, and the $d^4$ $Cr_3C_2$ and $Mo_3C_2$ materials for the capture of $H_2O$, predict spontaneous Gibbs free binding energies, but always lower than the ones due to the $CO_2$ chemisorption (see **Fig. 2**). In other words, the selected materials are more selective to interact with $CO_2$ than with $H_2O$, offering promising perspectives for the use of these materials in a water environment.

Concerning the nature of the interactions when bonding $CO_2$ on the MXenes surfaces, the low-coordinated metals atoms are seen to interact through the C of $CO_2$, potentially *via* electron-donation from the carbides (proximal distances between 2.17 and 2.49 Å), and also synergistically supported through intense interactions between the O lone pairs from $CO_2$ to the aforementioned three-fold metal atoms (proximal distances between 1.92 and 2.34 Å). In the case of the chemisorbed $H_2O$ molecules (**Fig. 2**), they bind through their O lone pairs to the metal atoms of the MXenes, showing also strong interactions, with $\Delta G_b$ between –0.21 and –0.64 eV. This is experimentally corroborated by the measured hydrophilic behavior of MXenes.[25]

The role of selectivity is crucial and strongly depends on how the material catalyzes the successive elementary electro-reductions. In this regard, test calculations employing classical DFT including dispersion *via* the DFT-D2 method (see **Table 1**) predicts non-spontaneous Gibbs free reaction energies at room temperature (hereafter simply referred as reaction energies) for the first H$^+$/e$^-$ pair gain, that leads to the chemisorbed HOCO• and OCHO• radical species. However, the hydrogenation on the C atom of the captured $CO_2$ molecule is more thermodynamically preferred than when acting on one of the two terminal O atoms of $CO_2$. This implies that the competitive CO and HCOOH formation would be minimal compared to •OCH$_2$O• formation, which follows with the third H$^+$/e$^-$ transfer onto one of the still unreacted O atoms to obtain the HOCH$_2$O• intermediate species. As result of the fourth H$^+$/e$^-$ gain, H$_2$CO is produced (with the release of one H$_2$O molecule), notwithstanding it deserves to be mentioned that the release of this captured H$_2$CO requires the input of a large amount of energy as a consequence of the aforementioned strong interactions between substrate and surface.

With the exception of Hf$_3$C$_2$, the formation of CH$_3$O• as a fifth-reduced radical species, which is the precursor of CH$_3$OH, exhibits spontaneous reaction energies, however the sixth hydrogenation at the CH$_3$O• radical requires its release from the surface in order that the H$^+$/e$^-$ pair can physically access to it, leading to the appearance of energy limitations in the range of 3 eV or higher for $d^2$ and Ta$_3$C$_2$ MXenes, and around 2.5 eV for M$_3$C$_2$ when M = V, Nb, Cr, and Mo. This limiting step eliminates this path, as well as the path towards the formation of CH$_4$ that also requires the injection of an important amount of energy for the release of the seventh hydrogenated CH$_3$• radical. However, the reactive nature of the transition metal carbides establishes an alternative path, that like in the case of the theoretical model of highly-reactive beryllium-doped BNs,[19] is thermodynamically preferred in the mechanism of CO$_2$ electroreduction to CH$_4$. Such a mechanism involves the sixth H$^+$/e$^-$ pair gain on the CH$_3$O• radical taking place on the CH$_3$ moiety, leading first to the release of CH$_4$ and an O atom inserted on the material, and secondly, continuing with two successive hydrogenations that produce an –OH• doped solid and a captured H$_2$O molecule. Despite such H$_2$O product contaminating the material, relative small reaction energy (with respect the limiting **OH•/**H$_2$O step) is needed, and besides that potential new CO$_2$ molecules will thermodynamically displace the captured water.

Although our DFT predictions establish a common path being exhibited by all the M$_3$C$_2$ MXenes studied in the present work, it is clear that $d^2$ Ti$_3$C$_2$, Zr$_3$C$_2$, and Hf$_3$C$_2$, as well as $d^3$ Ta$_3$C$_2$, are not promising for the CO$_2$ conversion into hydrocarbon fuels due to the large limiting steps predicted. For this reason more accurate DFT+$U$ calculations with explicit dispersion corrections *via* the DFT-D3 method have been carried out for the most promising $d^3$ V$_3$C$_2$ and Nb$_3$C$_2$, and $d^4$ and Cr$_3$C$_2$ and Mo$_3$C$_2$ materials.

Thus, and as previously mentioned, the inclusion of the Hubbard-like parameters *via* the DFT+$U$ approach to treat the strong on-site Coulomb interactions of the $d$-like localized electrons in M$_3$C$_2$ MXenes also predict spontaneous chemisorption processes of CO$_2$, however, values strongly differ due to the over-estimated interactions between CO$_2$ and the surface of the MXenes. Values of –0.77, –0.67, –0.44, and –1.01 eV are obtained for M = V, Nb, Cr, and Mo, respectively. Also, as a common feature, spontaneous reaction energies are predicted for the first $H^+/e^-$ gain to reach the OCHO• intermediate species, being –0.55, –0.38, –0.67, and –0.41 eV, respectively. These results for this first electron-reduction process are of paramount importance, highlighting a novel and impressive outcome in the spontaneous production of the OCHO• radical, or in other words, overcoming the limitation imposed by this classically limiting step.[18]

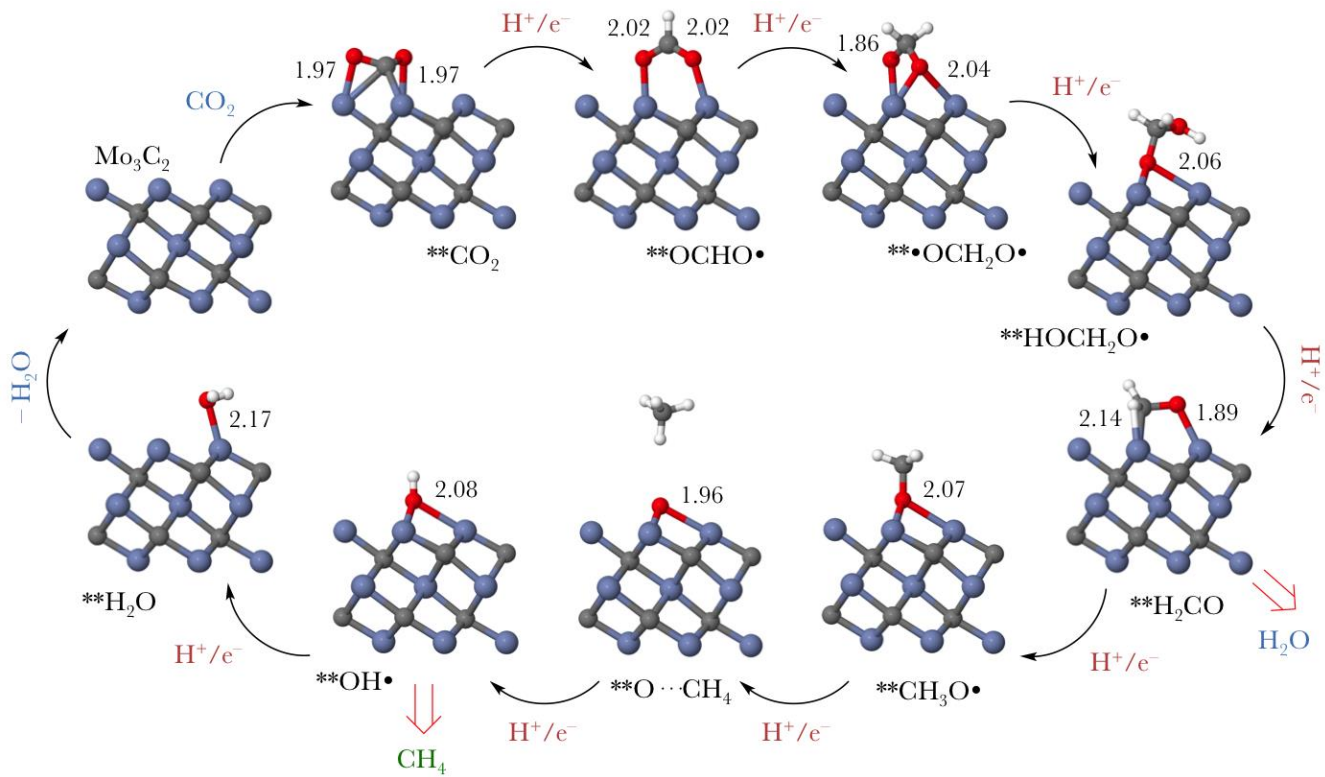

**Figure 3.** Minimum energy path (DFT+$U$/DFT-D3 calculations, $U$ = 3.5 eV) followed for the $CO_2$ conversion mechanism into *$CH_4$ and **$H_2O$ catalyzed by $Cr_3C_2$. Note: Grey, lilac, red, and white spheres refer to C, Cr, O, and H atoms, in that order. Note: * and ** symbols refer to physisorbed and chemisorbed species, respectively. Selected distances are indicated in Å.

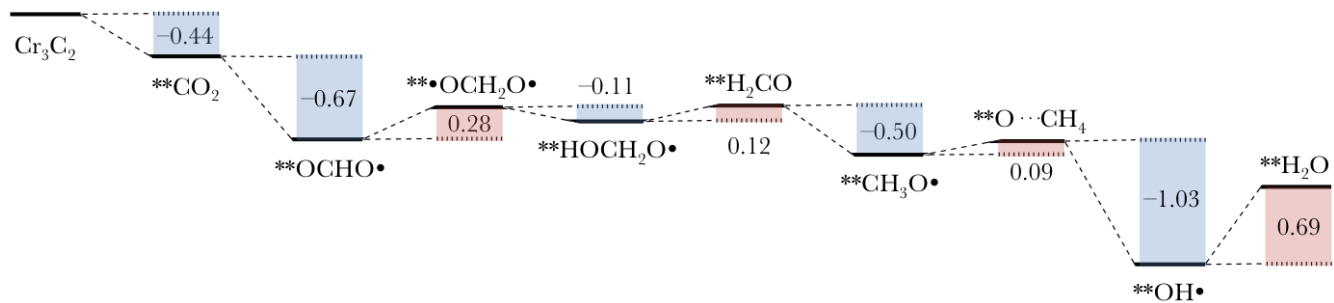

**Figure 4.** Gibbs free reaction energy diagram at 298.15 K (thermodynamics, in eV, referred to the immediately step) corresponding with the minimum energy path at the DFT+$U$/DFT-D3 computational level for the $CO_2$ conversion mechanism into *$CH_4$ and **$H_2O$ catalyzed by $Cr_3C_2$ ($U$ = 3.5 eV). Note: * and ** symbols refer to physisorbed and chemisorbed species, respectively.

With the exceptions of $Nb_3C_2$ and $Cr_3C_2$, a cascade of spontaneous elementary reactions occur for the second and third hydrogenations to obtain the chemisorbed •$OCH_2O$• and $HOCH_2O$• intermediate species. Small reaction energy of 0.28 eV in the second $H^+/e^-$ transfer is required for $Cr_3C_2$, while $HOCH_2O$• demands 0.80 eV for $Nb_3C_2$. Also, the $d^3$ MXenes show a different pattern in the production of $H_2CO$, which results in a highly distorted geometry with respect to the $sp^2$ hybridization as consequence of the strong interactions on the surface, This highlights the existence of a process that presents spontaneous reaction energies of –0.34 and –0.30 eV for M = V and Nb, whereas for the $d^4$ series (M = Cr and Mo) very low injections of energy, 0.12 and 0.06 eV, are needed. Furthermore, the $CH_3O$• is spontaneously pro-duced in all these cases. However, it deserves to be mentioned that, with the exception of the 0.80 eV that $Nb_3C_2$ requires for the production of $HOCH_2O\bullet$, the $CO_2$ conversion mechanism up to the fifth step follows a smooth path mainly characterized by spontaneous $H^+/e^-$ additions and their associated structural rearrangements.

As has been previously hypothesized, the $CO_2$ conversion mechanism catalyzed by $d^2$-$d^4$ $M_3C_2$ MXenes follows a common route in which the minimum energy path involves successive hydrogenations on the C and O atoms to reach $OCHO\bullet$, $\bullet OCH_2O\bullet$, and $HOCH_2O\bullet$, and obtaining $H_2CO$ as fourth-order reduced species in the form of a highly distorted and captured product. Since the chemisorbed $CH_3O\bullet$ radical is more thermodynamically preferred than the chemisorbed $H_2COH\bullet$ one during the fifth $H^+/e^-$ pair gain, our DFT+$U$/DFT-D3 results also predict that the classical route towards the formation of the $CH_3OH$ final product is not favored with respect to the sixth electro-reduction on the $CH_3$ moiety of the $CH_3O\bullet$ radical to reach a released $CH_4$ molecule and being an O atom covalently bounded on the material. As happened in the $HOCH_2O\bullet$ to $H_2CO$ step, $d^3$ MXenes are characterized to produce such $**O\cdot\cdot CH_4$ in a spontaneous process, while for $Cr_3C_2$ and $Mo_3C_2$, the injection of 0.09 and 0.35 eV are required.

Finally, the O-doped moiety of the surface is presented as highly reactive, being proof of this the highly spontaneous reactions consisting in the seventh hydrogenation to reach the $**OH\bullet$ radical species, that with exception of $Nb_3C_2$, indicates the maximum release of energy of all the elementary reactions (even more than the earlier $CO_2$ capture process). Unlike the classical materials in which the classical limiting step is imposed by the first hydrogenation step, our proposed materials present dramatically different behaviors, i.e. the release of such $**OH\bullet$ radical species in the form of a relatively intense chemisorbed $H_2O$ molecule once the eighth and final $H^+/e^-$ transfer comes into play. In this regard, $d^3$ MXenes exhibit values of 1.55, 1.75, and 1.24 eV for M = V, Nb, and Mo, which are relatively high with respect to the impressive catalytic performance shown by $Cr_3C_2$, which, as shown in **Figs. 3** and **4**, is hypothesized as the best alternative for the catalytic reduction of $CO_2$ based on transition metal carbide catalysis at the present time. For comparative purposes, the limiting step of the $Cr_3C_2$ outcomes are in the other of some tested and/or theoretically studied materials as for instance: Cu surface (Peterson *et al.*, $CO_2$ to $CH_4$, limiting step of 740 mV),[20] or graphene-supported amorphous $MoS_2$ (Li *et al.*, $CO_2$ to CO, over-potential of 540 mV when acting at a maximum faradaic efficiency).[64]

## SUMMARY AND CONCLUSIONS

In summary, our theoretical calculations predict that 2D $d^2$, $d^3$, and $d^4$ transition metal carbides (MXenes) with formulae $M_{n+1}C_n$ ($n = 2$) are capable of catalyzing $CO_2$ conversion into hydrocarbon compounds, being selective towards the formation of $CH_4$. Attending to the reaction energies computed in the present work by DFT modeling, $d^3$ $V_3C_2$ and $Nb_3C_2$, and $d^4$ and $Cr_3C_2$ and $Mo_3C_2$ materials offer the most promising results with limiting Gibbs free reaction energies at 298.15 K of 1.55, 1.75, 0.69, and 1.24 eV, respectively, according to the outcomes obtained at DFT+$U$ computational level plus explicit DFT-D3 dispersion corrections. Although the highly reactive behaviors of these set of Group IVB carbides (in addition to $d^3$ $Ta_3C_2$), the selected MXenes demonstrate active behaviors for the capture of $CO_2$, specially for those from the $d^2$ series (M = Ti, Zr, and Hf), dismissing their applicability in the conversion process. An analysis in-depth of the $CO_2$ conversion mechanism indicates that the minimum energy path follows an alternative route leading to the release of $CH_4$ during the sixth step, prior to the final $H_2O$ production. Important features have been predicted during the first hydrogenation step towards the formation of the $OCHO\bullet$ radical species, offering spontaneous energies for the promising MXenes. This point is of considerable significance since the first electro-reduction classically creates the severest obstacle of the entire reaction. Moreover, it is hypothesized that the capture of $CO_2$ during the early step of the process is spontaneous without passing through a physisorbed state. The strength of such capture is even larger than the computed binding energies for the chemisorption of $H_2O$. Against this background, we provide novel insights into the computer-aided catalyst design and the reaction mechanism modeling of the $CO_2$ conversion catalysts. We offer encouraging perspectives for the experimental testing of these materials in a water environment, profiling $Cr_3C_2$ as the best alternative of these series of MXenes with a limiting reaction energy of 0.69 eV.

## ASSOCIATED CONTENT

**Supporting Information**. Supporting Information contains: additional computational details, Gibbs free energies, density of states (HSE06 functional),[65-66] and Cartesian coordinates allowing an unambiguous reproducibility of the theoretical outcomes discussed in the present work. This material is available free of charge *via* the Internet at http://pubs.acs.org.

## AUTHORS INFORMATION

**Corresponding Authors**

*CS. Tel: (+61) 3 9902 9916. Fax: (+61) 3 9905 4597. E-mail: Chenghua.Sun@monash.edu

**Author Contributions**

All authors have given approval to the final version of the manuscript. ‡NL and LMA have equally contributed to the development of this work.

**Notes**

Authors declare no competing financial interests.

## ACKNOWLEDGMENTS

NL acknowledges the National Natural Science Foundation of China (Grant Nos. 51461135004, 51402225), the Doctoral Fund of Ministry of Education Priority Development Projects (Nos. 20130143130002, 20150303001), the Natural Science Foundation (NSF) of Hubei Province (No. 2015CFB227), the Key Technology Innovation Project of Hubei Province (No. 2013AAA005), the Fundamental Research Funds for the Central Universities (WUT: 2015-IVA-051), and the research board of the State Key Laboratory of Silicate Materials for Architectures (No. 47152005). LMA, DRM, and CS acknowledge the Australian Research Council (ARC) for its support through the ARC Centre of Electromaterials Science (ACES), Discover Project (DP130100268, CS), Future Fellowship (FT130100076, CS), and Laureate Fellow (DRM) schemes. We also thank the National Energy Research Scientific Computing Center in Shanghai and the National Computational Infrastructure (NCI), which is supported by the Australian Government, for providing the computational resources.

## REFERENCES

(1) Maginn, E. J. What to Do with $CO_2$. *J. Phys. Chem. Lett.* **2010**, *1*, 3478-3479.

(2) Karl, T. R.; Trenberth, K. E. Modern Global Climate Change. *Science* **2003**, *302*, 1719-1723.

(3) Kondratenko, E. V.; Mul, G.; Baltrusaitis, J.; Larrazábal, G. O.; Pérez-Ramírez, J. Status and Perspectives of $CO_2$ Conversion into Fuels and Chemicals by Catalytic, Photocatalytic and Electrocatalytic Processes. *Energy Environ. Sci.* **2013**, *6*, 3112-3135.

(4) Varghese, O. K.; Paulose, M.; LaTempa, T. J.; Grimes, C. A. High-Rate Solar Photocatalytic Conversion of $CO_2$ and Water Vapor to Hydrocarbon Fuels. *Nano Lett.* **2009**, *9*, 731-737.

(5) Liu, G.; Hoivik, N.; Wang, K.; Jakobsen, H. Engineering $TiO_2$ Nanomaterials for $CO_2$ Conversion/Solar Fuels. *Sol. Energy Mater. Sol. Cells* **2012**, *105*, 53-68.

(6) Zhao, C.; Liu, L.; Zhang, Q.; Wang, J.; Li, Y. Photocatalytic Conversion of $CO_2$ and $H_2O$ to Fuels by Nanostructured Ce–$TiO_2$/SBA-15 Composites. *Catal. Sci. Tech.* **2012**, *2*, 2558-2568.

(7) Graciani, J.; Mudiyanselage, K.; Xu, F.; Baber, A. E.; Evans, J.; Senanayake, S. D.; Stacchiola, D. J.; Liu, P.; Hrbek, J.; Sanz, J. F.; Rodriguez, J. A. Highly Active Copper-Ceria and Copper-Ceria-Titania Catalysts for Methanol Synthesis from $CO_2$. *Science* **2014**, *345*, 546-550.

(8) Lim, D.-H.; Jo, J. H.; Shin, D. Y.; Wilcox, J.; Ham, H. C.; Nam, S. W. Carbon Dioxide Conversion into Hydrocarbon Fuels on Defective Graphene-Supported Cu Nanoparticles from First Principles. *Nanoscale* **2014**, *6*, 5087-5092.

(9) Li, H.; Zhang, X.; MacFarlane, D. R. Carbon Quantum Dots/$Cu_2O$ Heterostructures for Solar-Light-Driven Conversion of $CO_2$ to Methanol. *Adv. Energy Mater.* **2015**, *5*, 1401077.

(10) Hsu, H.-C.; Shown, I.; Wei, H.-Y.; Chang, Y.-C.; Du, H.-Y.; Lin, Y.-G.; Tseng, C.-A.; Wang, C.-H.; Chen, L.-C.; Lin, Y.-C.; Chen, K.-H. Graphene Oxide as a Promising Photocatalyst for $CO_2$ to Methanol Conversion. *Nanoscale* **2013**, *5*, 262-268.

(11) Mao, J.; Peng, T.; Zhang, X.; Li, K.; Ye, L.; Zan, L. Effect of Graphitic Carbon Nitride Microstructures on the Activity and Selectivity of Photocatalytic $CO_2$ Reduction under Visible Light. *Catal. Sci. Tech.* **2013**, *3*, 1253-1260.

(12) Rosen, B. A.; Salehi-Khojin, A.; Thorson, M. R.; Zhu, W.; Whipple, D. T.; Kenis, P. J. A.; Masel, R. I. Ionic Liquid–Mediated Selective Conversion of $CO_2$ to CO at Low Overpotentials. *Science* **2011**, *334*, 643-644.

(13) DiMeglio, J. L.; Rosenthal, J. Selective Conversion of $CO_2$ to CO with High Efficiency Using an Inexpensive Bismuth-Based Electrocatalyst. *J. Am. Chem. Soc.* **2013**, *135*, 8798-8801.

(14) Li, P.; Jing, H.; Xu, J.; Wu, C.; Peng, H.; Lu, J.; Lu, F. High-Efficiency Synergistic Conversion of $CO_2$ to Methanol using $Fe_2O_3$ Nanotubes Modified with Double-Layer $Cu_2O$ Spheres. *Nanoscale* **2014**, *6*, 11380-11386.

(15) He, Y.; Wang, Y.; Zhang, L.; Teng, B.; Fan, M. High-Efficiency Conversion of $CO_2$ to Fuel over ZnO/g-$C_3N_4$ Photocatalyst. *Appl. Catal. B Environ.* **2015**, *168–169*, 1-8.

(16) Whipple, D. T.; Kenis, P. J. A. Prospects of $CO_2$ Utilization via Direct Heterogeneous Electrochemical Reduction. *J. Phys. Chem. Lett.* **2010**, *1*, 3451-3458.

(17) Habisreutinger, S. N.; Schmidt-Mende, L.; Stolarczyk, J. K. Photocatalytic Reduction of $CO_2$ on $TiO_2$ and Other Semiconductors. *Angew. Chem. Int. Ed.* **2013**, *52*, 7372-7408.

(18) Koppenol, W. H.; Rush, J. D. Reduction Potential of the Carbon Dioxide/Carbon Dioxide Radical Anion: A Comparison with other C1 Radicals. *J. Phys. Chem.* **1987**, *91*, 4429-4430.

(19) Azofra, L. M.; MacFarlane, D. R.; Sun, C. And Intensified $\pi$-Hole in Beryllium-Doped Boron Nitride Meshes: Its Determinant Role in the $CO_2$ Conversion into Hydrocarbon Fuels. *Chem. Commun.* **2016**, *52*, 3548-3551.

(20) Peterson, A. A.; Abild-Pedersen, F.; Studt, F.; Rossmeisl, J.; Nørskov, J. K. How Copper Catalyzes the Electroreduction of Carbon Dioxide into Hydrocarbon Fuels. *Energy Environ. Sci.* **2010**, *3*, 1311-1315.

(21) Peterson, A. A.; Nørskov, J. K. Activity Descriptors for $CO_2$ Electroreduction to Methane on Transition-Metal Catalysts. *J. Phys. Chem. Lett.* **2012**, *3*, 251-258.

(22) Lang, L. Beryllium: A Chronic Problem. *Environ. Health Perspect.* **1994**, *102*, 526-531.

(23) Naguib, M.; Kurtoglu, M.; Presser, V.; Lu, J.; Niu, J.; Heon, M.; Hultman, L.; Gogotsi, Y.; Barsoum, M. W. Two-Dimensional Nanocrystals Produced by Exfoliation of $Ti_3AlC_2$. *Adv. Mater.* **2011**, *23*, 4248-4253.

(24) Naguib, M.; Mochalin, V. N.; Barsoum, M. W.; Gogotsi, Y. 25[th] Anniversary Article: MXenes: A New Family of Two-Dimensional Materials. *Adv. Mater.* **2014**, *26*, 992-1005.

(25) Naguib, M.; Mashtalir, O.; Carle, J.; Presser, V.; Lu, J.; Hultman, L.; Gogotsi, Y.; Barsoum, M. W. Two-Dimensional Transition Metal Carbides. *ACS Nano* **2012**, *6*, 1322-1331.

(26) Naguib, M.; Halim, J.; Lu, J.; Cook, K. M.; Hultman, L.; Gogotsi, Y.; Barsoum, M. W. New Two-Dimensional Niobium and Vanadium Carbides as Promising Materials for Li-Ion Batteries. *J. Am. Chem. Soc.* **2013**, *135*, 15966-15969.

(27) Yang, J.; Naguib, M.; Ghidiu, M.; Pan, L.-M.; Gu, J.; Nanda, J.; Halim, J.; Gogotsi, Y.; Barsoum, M. W. Two-Dimensional Nb-Based $M_4C_3$ Solid Solutions (MXenes). *J. Am. Ceram. Soc.* **2015**, in press. DOI: 10.1111/jace.13922

(28) Anasori, B.; Xie, Y.; Beidaghi, M.; Lu, J.; Hosler, B. C.; Hultman, L.; Kent, P. R. C.; Gogotsi, Y.; Barsoum, M. W. Two-Dimensional, Ordered, Double Transition Metals Carbides (MXenes). *ACS Nano* **2015**, *9*, 9507-9516.

(29) Naguib, M.; Come, J.; Dyatkin, B.; Presser, V.; Taberna, P.-L.; Simon, P.; Barsoum, M. W.; Gogotsi, Y. MXene: A Promising Transition Metal Carbide Anode for Lithium-Ion Batteries. *Electrochem. Commun.* **2012**, *16*, 61-64.

(30) Xie, Y.; Naguib, M.; Mochalin, V. N.; Barsoum, M. W.; Gogotsi, Y.; Yu, X.; Nam, K.-W.; Yang, X.-Q.; Kolesnikov, A. I.; Kent, P. R. C. Role of Surface Structure on Li-Ion Energy Storage Capacity of Two-Dimensional Transition-Metal Carbides. *J. Am. Chem. Soc.* **2014**, *136*, 6385-6394.

(31) Liang, X.; Garsuch, A.; Nazar, L. F. Sulfur Cathodes Based on Conductive MXene Nanosheets for High-Performance Lithium–Sulfur Batteries. *Angew. Chem. Int. Ed.* **2015**, *54*, 3907-3911.

(32) Er, D.; Li, J.; Naguib, M.; Gogotsi, Y.; Shenoy, V. B. $Ti_3C_2$ MXene as a High Capacity Electrode Material for Metal (Li, Na, K, Ca) Ion Batteries. *ACS Appl. Mater. Interfaces* **2014**, *6*, 11173-11179.

(33) Lukatskaya, M. R.; Mashtalir, O.; Ren, C. E.; Dall'Agnese, Y.; Rozier, P.; Taberna, P. L.; Naguib, M.; Simon, P.; Barsoum, M. W.; Gogotsi, Y. Cation Intercalation and High Volumetric Capacitance of Two-Dimensional Titanium Carbide. *Science* **2013**, *341*, 1502-1505.

(34) Ghidiu, M.; Lukatskaya, M. R.; Zhao, M.-Q.; Gogotsi, Y.; Barsoum, M. W. Conductive Two-Dimensional Titanium Carbide 'Clay' with High Volumetric Capacitance. *Nature* **2014**, *516*, 78-81.

(35) Wang, X.; Kajiyama, S.; Iinuma, H.; Hosono, E.; Oro, S.; Moriguchi, I.; Okubo, M.; Yamada, A. Pseudocapacitance of MXene

Nanosheets for High-Power Sodium-Ion Hybrid Capacitors. *Nat. Commun.* **2015**, *6*, 6544.

(36) Xie, X.; Chen, S.; Ding, W.; Nie, Y.; Wei, Z. An Extraordinarily Stable Catalyst: Pt NPs supported on Two-Dimensional $Ti_3C_2X_2$ (X = OH, F) Nanosheets for Oxygen Reduction Reaction. *Chem. Commun.* **2013**, *49*, 10112-10114.

(37) Mashtalir, O.; Cook, K. M.; Mochalin, V. N.; Crowe, M.; Barsoum, M. W.; Gogotsi, Y. Dye Adsorption and Decomposition on Two-Dimensional Titanium Carbide in Aqueous Media. *J. Mater. Chem. A* **2014**, *2*, 14334-14338.

(38) Chen, W.-F.; Muckerman, J. T.; Fujita, E. Recent Developments in Transition Metal Carbides and Nitrides as Hydrogen Evolution Electrocatalysts. *Chem. Commun.* **2013**, *49*, 8896-8909.

(39) Liu, H.; Duan, C.; Yang, C.; Shen, W.; Wang, F.; Zhu, Z. A Novel Nitrite Biosensor based on the Direct Electrochemistry of Hemoglobin Immobilized on MXene-$Ti_3C_2$. *Sens. Actuators, B* **2015**, *218*, 60-66.

(40) Halim, J.; Lukatskaya, M. R.; Cook, K. M.; Lu, J.; Smith, C. R.; Näslund, L.-Å.; May, S. J.; Hultman, L.; Gogotsi, Y.; Eklund, P.; Barsoum, M. W. Transparent Conductive Two-Dimensional Titanium Carbide Epitaxial Thin Films. *Chem. Mater.* **2014**, *26*, 2374-2381.

(41) Furimsky, E. Metal Carbides and Nitrides as Potential Catalysts for Hydroprocessing. *Appl. Catal., A* **2003**, *240*, 1-28.

(42) Peng, Q.; Guo, J.; Zhang, Q.; Xiang, J.; Liu, B.; Zhou, A.; Liu, R.; Tian, Y. Unique Lead Adsorption Behavior of Activated Hydroxyl Group in Two-Dimensional Titanium Carbide. *J. Am. Chem. Soc.* **2014**, *136*, 4113-4116.

(43) Liu, Y.; Kelly, T. G.; Chen, J. G.; Mustain, W. E. Metal Carbides as Alternative Electrocatalyst Supports. *ACS Catal.* **2013**, *3*, 1184-1194.

(44) Chen, W. F.; Wang, C. H.; Sasaki, K.; Marinkovic, N.; Xu, W.; Muckerman, J. T.; Zhu, Y.; Adzic, R. R. Highly Active and Durable Nanostructured Molybdenum Carbide Electrocatalysts for Hydrogen Production. *Energy Environ. Sci.* **2013**, *6*, 943-951.

(45) Lee, J. S. Metal Carbides, *Encyclopedia of Catalysis*, John Wiley & Sons, Inc., 2002.

(46) Perdew, J. P.; Burke, K.; Ernzerhof, M. Generalized Gradient Approximation Made Simple. *Phys. Rev. Lett.* **1996**, *77*, 3865-3868.

(47) Blöchl, P. E. Projector Augmented-Wave Method. *Phys. Rev. B* **1994**, *50*, 17953-17979.

(48) Kresse, G.; Joubert, D. From Ultrasoft Pseudopotentials to the Projector Augmented-Wave Method. *Phys. Rev. B* **1999**, *59*, 1758-1775.

(49) Grimme, S. Semiempirical GGA-Type Density Functional Constructed with a Long-Range Dispersion Correction. *J. Comput. Chem.* **2006**, *27*, 1787-1799.

(50) Liechtenstein, A. I.; Anisimov, V. I.; Zaanen, J. Density-Functional Theory and Strong Interactions: Orbital Ordering in Mott-Hubbard Insulators. *Phys. Rev. B* **1995**, *52*, R5467-R5470.

(51) Dudarev, S. L.; Botton, G. A.; Savrasov, S. Y.; Humphreys, C. J.; Sutton, A. P. Electron-Energy-Loss Spectra and the Structural Stability of Nickel Oxide: An LSDA+*U* Study. *Phys. Rev. B* **1998**, *57*, 1505-1509.

(52) Finazzi, E.; Di Valentin, C.; Pacchioni, G.; Selloni, A. Excess Electron States in Reduced Bulk Anatase $TiO_2$: Comparison of Standard GGA, GGA+*U*, and Hybrid DFT Calculations. *J. Chem. Phys.* **2008**, *129*, 154113.

(53) Jain, A.; Hautier, G.; Ong, S. P.; Moore, C. J.; Fischer, C. C.; Persson, K. A.; Ceder, G. Formation Enthalpies by Mixing GGA and GGA+*U* Calculations. *Phys. Rev. B* **2011**, *84*, 045115.

(54) Stevanović, V.; Lany, S.; Zhang, X.; Zunger, A. Correcting Density Functional Theory for Accurate Predictions of Compound Enthalpies of Formation: Fitted Elemental-Phase Reference Energies. *Phys. Rev. B* **2012**, *85*, 115104.

(55) Sun, C.; Liao, T.; Lu, G. Q.; Smith, S. C. The Role of Atomic Vacancy on Water Dissociation over Titanium Dioxide Nanosheet: A Density Functional Theory Study. *J. Phys. Chem. C* **2012**, *116*, 2477-2482.

(56) Grimme, S.; Antony, J.; Ehrlich, S.; Krieg, H. A Consistent and Accurate *Ab Initio* Parametrization of Density Functional Dispersion Correction (DFT-D) for the 94 Elements H-Pu. *J. Chem. Phys.* **2010**, *132*, 154104.

(57) Grimme, S.; Ehrlich, S.; Goerigk, L. Effect of the Damping Function in Dispersion Corrected Density Functional Theory. *J. Comput. Chem.* **2011**, *32*, 1456-1465.

(58) Kresse, G.; Hafner, J. *Ab Initio* Molecular Dynamics for Liquid Metals. *Phys. Rev. B* **1993**, *47*, 558-561.

(59) Kresse, G.; Hafner, J. *Ab Initio* Molecular-Dynamics Simulation of the Liquid-Metal–Amorphous-Semiconductor Transition in Germanium. *Phys. Rev. B* **1994**, *49*, 14251-14269.

(60) Kresse, G.; Furthmüller, J. Efficient Iterative Schemes for *Ab Initio* Total-Energy Calculations using a Plane-Wave Basis Set. *Phys. Rev. B* **1996**, *54*, 11169-11186.

(61) Kresse, G.; Furthmüller, J. Efficiency of *Ab-Initio* Total Energy Calculations for Metals and Semiconductors using a Plane-Wave Basis Set. *Comput. Mat. Sci.* **1996**, *6*, 15-50.

(62) Yin, W.-J.; Krack, M.; Wen, B.; Ma, S.-Y.; Liu, L.-M. $CO_2$ Capture and Conversion on Rutile $TiO_2$(110) in the Water Environment: Insight by First-Principles Calculations. *J. Phys. Chem. Lett.* **2015**, *6*, 2538-2545.

(63) While preparing this manuscript, a similar research concerning the function of transition metal carbides as $CO_2$ capture marerials has appeared at: Kunkel, C.; Viñes, F.; Illas, F. Transition Metal Carbides as Novel Materials for $CO_2$ Capture, Storage, and Activation. *Energy Environ. Sci.,* **2016**, *9*, 141-144.

(64) Li, F.; Zhao, S.-F.; Chen, L.; Khan, A.; MacFarlane, D. R.; Zhang, J. Polyethylenimine Promoted Electrocatalytic Reduction of $CO_2$ to CO in Aqueous Medium by Graphene-Supported Amorphous Molybdenum Sulphide. *Energy Envioron. Sci.* **2016**, *9*, 216-223.

(65) Heyd, J.; Scuseria, G. E.; Ernzerhof, M. Hybrid Functionals based on a Screened Coulomb Potential. *J. Chem. Phys.* **2003**, *118*, 8207-8215. See *Erratum* at: Heyd, J.; Scuseria, G. E.; Ernzerhof, M. Hybrid Functionals based on a Screened Coulomb Potential. *J. Chem. Phys.* **2006**, *124*, 219906.

(66) Heyd, J.; Scuseria, G. E. Efficient Hybrid Density Functional Calculations in Solids: Assessment of the Heyd-Scuseria-Ernzerhof Screened Coulomb Hybrid Functional. *J. Chem. Phys.* **2004**, *121*, 1187-1192.

Table of Contents (TOC)

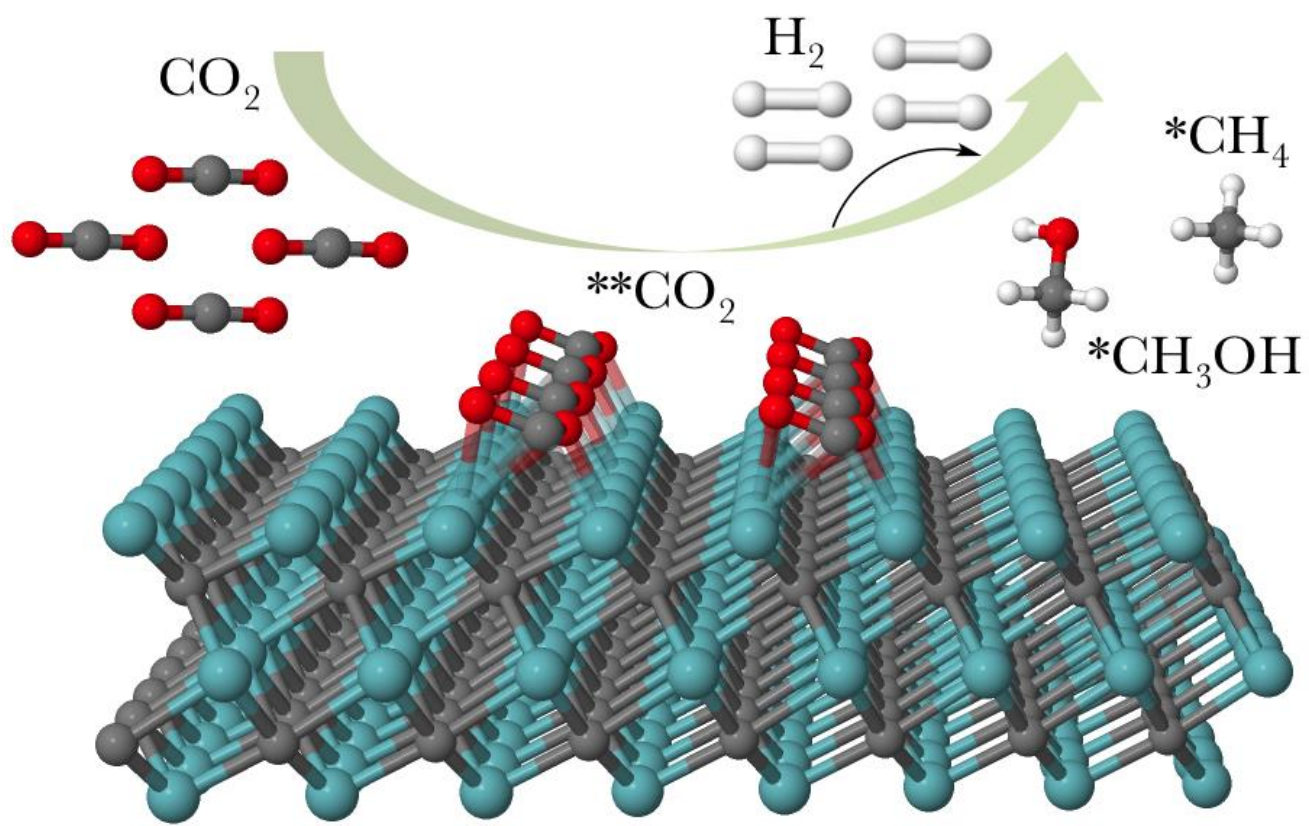